# Comment on "Is Faith the Enemy of Science" (to appear in "Physics in Canada")

Lawrence M. Krauss,
School of Earth and Space Exploration and Physics Department,
Arizona State University,
PO Box 871404, Tempe AZ 85287-1404

## Abstract

This comment was solicited by Physics in Canada and will appear alongside the article by Richard Mackenzie in the next issue.

When Richard Mackenzie contacted me some time ago asking for the slides of my presentation at the CAP annual meeting, I had no idea that he was planning such a comprehensive, and cogent, reflection on my remarks. After reading the substance of his paper [1], I find myself with little to disagree with. The chief disagreement we may have, if indeed we have one, is primarily semantic. It is based on the definition of the three key words, "faith", "ignorance" and "enemy".

Let me begin by explaining what I meant by the word "enemy". I take an operational view of this word. An enemy is someone to either be avoided or vanquished. I have had several interesting discussions with Richard Dawkins, some of them quite public and available on YouTube, on this issue. I have asked Richard if his recent purpose is to destroy faith or teach science, and he has indicated that destroying faith at the moment is a higher priority. I accept that argument, however for me the latter purpose, teaching science, is higher priority. (At least it certainly was in the context of a lecture on the teaching of science!) And since I therefore view that

vanquishing ignorance is a higher priority for a teacher, this makes ignorance the enemy

Now, let's talk about ignorance for a bit. There is nothing evil about ignorance. I always make a big point of stating that when I describe some viewpoint as being based on ignorance it is not a pejorative statement, but meant as a statement of fact. Thus, for example, when I say that President Bush's statement about evolution vs intelligent design that "Both sides should be taught, so students know what the debate is all about.", I argue that is a statement of ignorance, because he doesn't know there is no scientific debate at the current time. It is not a stupid statement. If there were such a debate, it would be worth teaching students about it. (I should add that I don't take a moralistic view of the term "enemy" either. Enemies need not be evil. They are simply enemies.)

Finally, the most emotionally charged word of all, "faith". Richard takes this to mean an unsubstantiated belief, which is not a bad definition. But he then interprets religious faith as being the same as faith in the precepts of organized religion. If this were true, I agree that science and religious faith are generally incompatible. There is nothing about the universe that science has unveiled that supports the notion of a God interested in human affairs, and many of the stories in the Bible, for example, are not empirically true. However, having a kind of general faith in order and purpose to the Universe is not so obviously unscientific, and while I don't view this faith as particularly well founded, I also don't view it as particularly destructive.

But, just for the purposes of discussion, what about doctrinal faith by religious scientists? Even if it is inconsistent with science, is this something that we need to vanquish? I doubt we can, and I don't see trying to do so as the highest priority.

First, I see the existence of conventionally religious scientists as merely a clear example of the fact that humans can hold fast to two inconsistent ideas at the same time. This is not a fact worth extolling, but it is simply something that we cannot do much about. Humans are not completely logical beings. As I have said elsewhere, most of us need to convince ourselves of 10 impossible things before breakfast in order to face the day. Perhaps the world would be a better place if human nature was completely logical, but it isn't. Should we therefore place our highest priority on vanquishing this aspect of human nature? I remain unconvinced.

Second, as long as someone's religious faith does not get in the way of their learning about nature, their ability to assess empirical data, and to predict the results of future experiments, then I view it as no more obstructionist than the faith they may have that money can't by happiness, or that marriage produces happiness ever after, or that the Canadiens will win the Stanley Cup. Naïve, perhaps. Maybe even based on ignorance. But not necessarily counterproductive.

[1] Richard Mackenzie, arxiv:0807.3670, to appear *Physics in Canada.*